6-20-10

# The parity operator in quantum optical metrology

*Christopher C. Gerry and Jihane Mimih*

*Department of Physics and Astronomy*

*Lehman College, The City University of New York*

*Bronx, New York 10468-1589*

*USA*

Photon number states are assigned a parity of $+1$ if their photon number is even and a parity of $-1$ if odd. The parity operator, which is minus one to the power of the photon number operator, is a Hermitian operator and thus a quantum mechanical observable though it has no classical analog, the concept being meaningless in the context of classical light waves. In this paper we review work on the application of the parity operator to the problem of quantum metrology for the detection of small phase shifts with quantum optical interferometry using highly entangled field states such as the so-called N00N states, and states obtained by injecting twin Fock states into a beam splitter. With such states and with the performance of parity measurements on one of the output beams of the interferometer, one can breach the standard quantum limit, or shot-noise limit, of sensitivity down to the Heisenberg limit, the greatest degree of phase sensitivity allowed by quantum mechanics for linear phase shifts. Heisenberg limit sensitivities are expected to eventually play an important role in attempts to detect gravitational waves in interferometric detection systems such as LIGO and VIRGO.

# I. Introduction

Mention the word "parity" in the company of physicists and most will likely think of the discrete symmetry operation which constitutes a change in the sign of all the spatial coordinates [1, 2]; that is, the transformation $(x, y, z) \rightarrow (-x, -y, -z)$, or $\mathbf{r} \rightarrow -\mathbf{r}$. This operation is also known as an "inversion." The strong and electromagnetic interactions are symmetric under this operation, but, as is well known, the weak interaction is not [1, 2]. On the other hand, in the context of elementary mathematics parity simply refers to whether an integer is even or odd. If we restrict ourselves to one dimension, wherein the parity operation is reduced to $x \rightarrow -x$, we can see a relation between the two definitions. For example, for the wave function $\psi_n(x)$ belonging to the $n$th energy level of a one-dimensional harmonic oscillator [3] we find that $\psi_n(-x) = (-1)^n \psi_n(x)$ so that $\psi_n(-x) = \psi_n(x)$ if $n$ is even and $\psi_n(-x) = -\psi_n(x)$ if $n$ is odd in accordance with what we mean by the notion of even and odd functions [4]. Of course, the parity of wave functions is responsible for the selection rules as revealed by spectroscopy [5].

Independent of the symmetry considerations just discussed, there is the purely mathematical notion of parity wherein the integers are assigned a parity of +1 for even or −1 for odd. That is, integer $l$ has a parity given by $\Pi = (-1)^l$. Obviously, the two notions of parity we have discussed are related as we can see for the harmonic oscillator wave functions discussed in the previous paragraph. But suppose we ignore configuration space (wave functions) and work in Hilbert space where our basis vectors are denoted by $\{|l\rangle, l \text{ an integer}\}$ and where $|l\rangle$ is an eigenket of some operator $\hat{l}$, that is $\hat{l}|l\rangle = l|l\rangle$. We can then define a *parity operator* within that space as $\hat{\Pi} = (-1)^{\hat{l}}$ such that $\hat{\Pi}|l\rangle = (-1)^l |l\rangle$. For example, for a one-dimensional harmonic oscillator the energy eigenkets are given by $\{|n\rangle, n = 0,1,2....\}$ such that for the Hamiltonian $\hat{H} = \hbar\omega(\hat{n} + \tfrac{1}{2})$, $\hat{H}|n\rangle = E_n |n\rangle$, where $E_n = \hbar\omega(n + \tfrac{1}{2})$, and where we have $\hat{n}|n\rangle = n|n\rangle$. Thus the parity operator for this system is $\hat{\Pi} = (-1)^{\hat{n}}$. As another example, for an angular momentum system of basis states $|j,m\rangle$ the corresponding parity operator is $\hat{\Pi} = (-1)^{j+\hat{J}_z/\hbar}$ where $\hat{J}_z |j,m\rangle = \hbar m |j,m\rangle$. One could just as well define the parity operator as $\hat{\Pi} = (-1)^{j-\hat{J}_z/\hbar}$ for this system.

Being Hermitian, the parity operator is a quantum mechanical observable, but, unlike other quantum observables, it has no classical counterpart. So is it useful? Yes! It turns out that the quasi-probability phase-space distribution known as the Wigner function [6] can be written as the expectation value of the displaced parity operator [7, 8]. It has also been used as a dichotomic variable in order to implement [9] the Clauser, Horne, Shimony, Holt [10] form of Bell's inequality [11]. In what follows we discuss the use of the parity operator in quantum optical metrology.

Quantum metrology [12] is the art and science of using quantum mechanical states of matter or light to perform measurements with quantum enhanced sensitivities, the enhancement being directly due to the non-classical natures of the states employed. By "quantum enhanced measurements" we mean measurements that lead to the enhanced estimation of physical parameters that becomes possible only because of the quantum mechanical properties of the medium used to extract the signal. The issues of quantum enhanced measurements should not be confused with the interpretational issues surrounding the quantum measurement problem, sometimes called the measurement paradox [13].

In this paper, we have in mind the detection of small phase shifts of the type expected in optical interferometer gravitational wave detectors, such as LIGO [14] and VIRGO detectors [15]. A popular account of the history of gravitational wave detection has recently been given by Bartusiak [16]. Optical approaches to the detection of these waves have been the driving force behind much of the effort towards using non-classical states of light, such as squeezed states, in order to reduce the quantum noise. Gravitational waves are expected to be so weak that their signal would ordinarily be lost within the quantum noise, i.e. the quantum mechanical fluctuations, of the detector. Hence the effort to reduce quantum noise (actually to redistribute it) such that in the measured quantity, a phase shift, the quantum noise in that quantity is reduced below the so-called quantum standard limit (SQL) [17, 18]. We can understand this using the heuristic number-phase uncertainty relation $\Delta N \Delta \varphi \geq 1$ where $\Delta N$ is the uncertainty in the number of photons. If the uncertainty relation is approximately equalized $\left(\Delta N \Delta \varphi \approx 1\right)$ then to reach greater sensitivity in the measurement of the phase shift, i.e. to reduce the fluctuations symbolized by $\Delta \varphi$, we are evidently required to have large fluctuations in the photon number. The smaller we can make $\Delta \varphi$, the greater the sensitivity of the measurements of the phase shift. And they will need to be very sensitive, indeed. Already, the LIGO detector is so sensitive that it can detect phase shifts corresponding to displacements of $10^{-18}$ m [19], a thousandth of the diameter of a proton. But that's not good enough for the detection of gravitational waves!

Here we describe the work we have done over the past decade [20-29] on the use of the optical parity operator and of the various quantum field states we have considered as candidates for approaching the highest level of sensitivity in the detection of small phase shifts via optical interferometry. Our parity operator approach derives itself from another area where quantum enhanced measurements have been pursued, namely in precision frequency measurements in experiments with trapped atoms or ions, of importance for frequency standards and atomic clocks. In 1996, Bollinger *et al*. [30] discussed the use of what amounts to the angular momentum parity operator of above for a system of trapped two-level ions, a pseudo-spin system, in which parity is determined via counting the number of ions of the sample that populate the excited state. Transition frequencies can be determined in a Ramsey-type spectroscopy experiment [31] with a sensitivity that goes as $\delta\omega \sim 1/NT$ where *N* is the total number of ions and *T* is the time between Ramsey pulses. Such a level of sensitivity is called Heisenberg-limited. Experiments of this type have been performed [32] with small numbers of ions. As the Ramsey spectroscopic method of separated fields is identical mathematically to optical interferometry it seemed sensible, as we shall see, to investigate the utility of the photon number parity operator in this context.

On the subject of quantum enhanced metrology there already appeared in recent years three relevant articles in Contemporary Physics; these are by Sergienko and Jaeger [33] in 2003, by Dunningham [34] in 2006, and by Dowling [35] in 2008. Those articles, particularly the Dowling one, could usefully be consulted while reading the present one.

The paper is organized as follows: In section II we review interferometry performed with classical states of light, the so-called coherent states. In section III we introduce the N00N states and show how they lead to Heisenberg-limited phase shift estimations if parity measurements are performed on one of the exit beams of the interferometer. In section IV we discuss the use of parity measurements in a somewhat different scenario wherein twin-Fock states are simultaneously injected into both inputs of the first beam splitter of an interferometer. In Section V we discuss the connection between the joint photon number probability distribution inside an interferometer with Heisenberg uncertainty measurements. In section VI we discuss the issue of measuring photon number parity, and in section VII we conclude the paper with some brief remarks. In the Appendix we discuss a simple method to obtain output beam splitter states for given input states.

**II. Interferometry with classical-like states: coherent states**

We begin with a brief lesson on the quantized electromagnetic field. The quantized electromagnetic field can be understood as a collection of harmonic oscillators where each of the infinite number of modes of the field is specified uniquely by its direction of propagation, frequency (energy), and polarization. For a single mode of the field with frequency $\omega$, propagation direction specified by the wave vector **k** (where $|\mathbf{k}| = 2\pi/\lambda$), and polarization **e**, the electric field operator can be represented as [36]

$$\hat{\mathbf{E}}(\mathbf{r},t) = i\left(\frac{\hbar\omega}{2\varepsilon_0 V}\right)^{1/2} \mathbf{e}\left[\hat{a}e^{i(\mathbf{k}\bullet\mathbf{r}-\omega t)} - \hat{a}^\dagger e^{-i(\mathbf{k}\bullet\mathbf{r}-\omega t)}\right] \qquad (1)$$

where $\hat{a}$ and $\hat{a}^\dagger$ are the usual annihilation and creation operators of a one-dimensional harmonic oscillator satisfying the commutation relation $[\hat{a},\hat{a}^\dagger] = 1$. The energy operator, or Hamiltonian, associated with this field mode is $\hat{H} = \hbar\omega(\hat{a}^\dagger\hat{a} + \frac{1}{2})$, which is, of course, identical to that of a one-dimensional harmonic oscillator as can be found in most undergraduate quantum mechanics textbooks. The number operator $\hat{n} \equiv \hat{a}^\dagger \hat{a}$ has eigenstates $|n\rangle$ such that $\hat{a}^\dagger \hat{a}|n\rangle = n|n\rangle$ yield the familiar oscillator energy eigenvalues $E_n = \hbar\omega(n+\tfrac{1}{2})$. In the context of quantum field theory, the number $n$ is understood to represent the number of photons occupying the mode. The number states are orthonormal, i.e. $\langle n'|n\rangle = \delta_{n',n}$, and the action of the creation and annihilation operators on the states is to raise and lower the number of photons by one: $\hat{a}^\dagger|n\rangle = \sqrt{n+1}|n+1\rangle$, $\hat{a}|n\rangle = \sqrt{n}|n\rangle$. Because of the orthogonality of number states with different numbers of photons, we have $\langle n|\hat{\mathbf{E}}(\mathbf{r},t)|n\rangle = 0$. This is true no matter how large the photon number $n$, thus demonstrating that the limit of large quantum numbers is not a classical limit. A classical radiation field should oscillate in time and space. A quantum state of the field with these classical-like properties is the Glauber [37] coherent state $|\alpha\rangle$ which satisfies the eigenvalue problems $\hat{a}|\alpha\rangle = \alpha|\alpha\rangle$, $\langle\alpha|\hat{a}^\dagger = \alpha^*\langle\alpha|$ where $\alpha$ is a complex number. That $\alpha$ can be complex is due to the fact that $\hat{a}$ is not a Hermitian operator. If we take the expectation value of the electric field operator with a coherent state we find

$$\langle\alpha|\hat{\mathbf{E}}(\mathbf{r},\mathrm{t})|\alpha\rangle = i\left(\frac{\hbar\omega}{2\varepsilon_0 V}\right)^{1/2} \mathbf{e}\left[\alpha e^{i(\mathbf{k}\bullet\mathbf{r}-\omega t)} - \alpha^* e^{-i(\mathbf{k}\bullet\mathbf{r}-\omega t)}\right] \qquad (2)$$

which is exactly of the form of a classical radiation field. In terms of the photon number states, the coherent states may be expressed as

$$|\alpha\rangle = e^{-|\alpha|^2/2}\sum_{n=0}^{\infty}\frac{\alpha^n}{\sqrt{n!}}|n\rangle. \qquad (3)$$

They are normalized, $\langle \alpha | \alpha \rangle = 1$, but not orthogonal as $\langle \beta | \alpha \rangle \neq 0$ for $\alpha$ and $\beta$ different. The average photon number in the field is $\bar{n} = \langle \alpha | \hat{n} | \alpha \rangle = \langle \alpha | \hat{a}^\dagger \hat{a} | \alpha \rangle = |\alpha|^2$, and the photon number distribution for such a field state is a Poisson distribution: $P(n) = |\langle n | \alpha \rangle|^2 = e^{-\bar{n}} \bar{n}^n / n!$. The importance of coherent states is that they represent the output light of a phase-stabilized laser. In the balance of this article, we'll assume that coherent light and laser light are synonymous. Even though the coherent state is very classical-like, it is, nevertheless, a quantum state and therefore does carry quantum noise, a manifestation of which we shall encounter shortly.

So now let's consider interferometry as it is performed with coherent states of light, i.e. with light from a phase-stabilized laser. A schematic of our interferometer is given in Fig. 1. Pictured is a so-called Mach-Zehnder interferometer (MZI) consisting of two beam splitters, which we assume to be 50:50, with two pathways from the first beam splitter to the second where the difference in the optical path lengths is represented by the phase shift $\varphi$. At the output of the second beam splitter are placed photo-detectors. As was pointed out by Dowling [35], an MZI is nothing more than an unfolded Michelson interferometer, so the operation of the former is essentially identical to that of the latter. We label the interferometer beams *a* and *b* for counter-clockwise and clockwise paths, respectively, as indicated in Fig. 1.

Before proceeding to interferometry, we need to discuss the quantum theory of a beam splitter. We consider only 50:50 beam splitters and ones where the reflected wave picks up a $\pi/2$ phase shift. As indicated in Fig. 2, we label the input beams as *a* and *b* and the output beams as $a'$ and $b'$. The corresponding Bose operators $(\hat{a},\hat{b})$ and $(\hat{a}',\hat{b}')$ are related according to the transformation

$$\hat{a}' = \frac{1}{\sqrt{2}}(\hat{a}+i\hat{b}), \quad \hat{b}' = \frac{1}{\sqrt{2}}(\hat{b}+i\hat{a}). \tag{4}$$

The reader can easily verify that the proper commutation relations are satisfied for these transformed operators, a necessary requirement for the preservation of unitarity across the beam splitter. By the way, a beam splitter does just what its name says: it splits beams; it does not split photons. In this sense it is a passive device.

We are interested, then, in the transformation of quantum states of light by a beam splitter, and we first examine what happens to coherent state inputs as we can use classical reasoning to obtain the correct output states. We take as inputs the coherent states $|\alpha\rangle_a$ and $|\beta\rangle_b$, that is, the product state $|\alpha\rangle_a |\beta\rangle_b$. Keeping in mind that our beam splitter imparts a $\pi/2$ phase shift to the reflected beam, i.e. that the reflected beam amplitude picks up the factor $i$, and that a 50:50 beam splitter splits the intensities of each of the incoming beams equally, the beam splitter causes the transformation

$$|\alpha\rangle_a |\beta\rangle_b \xrightarrow{\text{BS}} \left|\frac{\alpha+i\beta}{\sqrt{2}}\right\rangle_{a'} \left|\frac{\beta+i\alpha}{\sqrt{2}}\right\rangle_{b'}. \tag{5}$$

Note that the total average photon number of the two beams, $|\alpha|^2 + |\beta|^2$, is preserved across the beam splitter. In the special case where we have a vacuum state in the lower left beam, i.e. for $\beta = 0$, we have the transformation

$$|\alpha\rangle_a |0\rangle_b \xrightarrow{\text{BS}} \left|\frac{\alpha}{\sqrt{2}}\right\rangle_{a'} \left|\frac{i\alpha}{\sqrt{2}}\right\rangle_{b'}. \tag{6}$$

The results for the beam splitting of coherent states are exactly the same as those we'd expect for classical coherent light. In the Appendix we obtain this result a bit more rigorously, and we consider the transformation of other input states, such as the product number states of the form $|N\rangle_a |M\rangle_b$ as the results will be needed in later sections. Notice that for input coherent states, including the vacuum, the quantum states of the output beams are just a product of coherent states, one for each beam. A beam splitter does not create entanglement for input coherent states. That this is so is just another reason to consider the coherent state as the most classical state of all pure states of light. If a two-particle (or two-mode, in this case) state can be written as a product of the states for each particle or mode, the state is factorizable, i.e. is not entangled. If it is not possible to factorize a state in any basis, the state is said to be entangled. If two independent, hence unentangled, pure states of light enter the two inputs of a beam splitter, the output beams will be in an entangled state except for the very special case of input coherent states which does not create entanglement. In general, a beam splitter creates entanglement.

Henceforth we shall usually drop the primes in the labeling of the output beams, with the understanding that $a$ and $b$ label the counter-clockwise and clockwise beams, respectively of the MZI, as indicated in Fig. 1.

Now, let's go back to our interferometer. We take as input to the first beam splitter the states $|\alpha\rangle_a |0\rangle_b$, where we have used the labeling scheme mentioned above. Using the results of the previous paragraph, this beam splitter performs the transformation

$$|\alpha\rangle_a |0\rangle_b \xrightarrow{BS_1} \left|\frac{\alpha}{\sqrt{2}}\right\rangle_a \left|\frac{i\alpha}{\sqrt{2}}\right\rangle_b. \qquad (7)$$

The mirrors contribute $\pi/2$ phase shifts to both beams leading to an overall phase shift that can be ignored. The phase shift $\varphi$ in the clockwise beam is represented by the unitary operator $\hat{U}_{PS} = \exp(i\varphi\, \hat{n}_b)$, where $\hat{n}_b$ is the photon number operator associated with the clockwise beam, and where $\varphi = 2\pi x/\lambda$, $x$ being the difference in the lengths of the two paths inside the MZI. The phase shift is encoded as

$$\hat{U}_{PS} \left|\frac{\alpha}{\sqrt{2}}\right\rangle_a \left|\frac{i\alpha}{\sqrt{2}}\right\rangle_b = \left|\frac{\alpha}{\sqrt{2}}\right\rangle_a \left|\frac{i\alpha e^{i\varphi}}{\sqrt{2}}\right\rangle_b, \qquad (8)$$

where we have used the operation $\exp(i\varphi\, \hat{n}_b)|n\rangle_b = e^{in\varphi}|n\rangle_b$. At the second beam splitter we have, using the result of Eq. (5),

$$\left|\frac{\alpha}{\sqrt{2}}\right\rangle_a \left|\frac{i\alpha e^{i\varphi}}{\sqrt{2}}\right\rangle_b \xrightarrow{BS_2} \left|\frac{\alpha}{2}(1-e^{i\varphi})\right\rangle_a \left|\frac{i\alpha}{2}(1+e^{i\varphi})\right\rangle_b. \qquad (9)$$

The intensity $I$ of a quantized field is proportional to the average photon number of the quantum state of the field: $I \propto \langle \hat{n} \rangle$ [38] and thus the difference of the intensities at the outputs of the second beam splitter can be written as $\delta I \propto \langle \hat{J} \rangle$ where $\hat{J} = \hat{n}_b - \hat{n}_a$. We easily find that

$$\langle \hat{n}_a \rangle = \frac{\bar{n}}{2}[1-\cos\varphi], \quad \langle \hat{n}_b \rangle = \frac{\bar{n}}{2}[1+\cos\varphi], \qquad (10)$$

so that $\langle \hat{J} \rangle = \bar{n}\cos\varphi$, where $\bar{n} = |\alpha|^2$ is the average photon number of the input coherent state, and thus the phase shift $\varphi$ is detectable through its modulation of the difference in the output intensities of the MZI. In Fig. 3 we plot $\langle \hat{J} \rangle$ versus $\varphi$ for various $\bar{n}$. Note that the "frequency" of the oscillations depends only on $\varphi$ and not on the amplitude of the coherent field. The frequency of the oscillations is a measure of the resolution of the detection scheme.

In addition to detecting the phase-shift we need to know the precision, or sensitivity, of the measurement. From the error propagation calculus [39], the uncertainty in the estimation of the phase shift, $\Delta\varphi$, upon measurement of the operator $\hat{J}$ is given by

$$(\Delta\varphi)^2 = \frac{(\Delta J)^2}{\left(\partial\langle\hat{J}\rangle/\partial\varphi\right)^2}, \tag{11}$$

where $(\Delta J)^2 = \langle\hat{J}^2\rangle - \langle\hat{J}\rangle^2$. This estimate is valid for any input state to the interferometer as long as the operator $\hat{J}$ is the measured observable. As we desire $\Delta\varphi$ to be as small as possible we would like $\left|\partial\langle\hat{J}\rangle/\partial\varphi\right|$ as large as possible. For our input coherent state we obtain

$$\Delta\varphi = \frac{1}{\sqrt{\bar{n}}\,|\sin\varphi|}. \tag{12}$$

The minimum uncertainty that can be achieved is obviously $\Delta\varphi_{\min} = 1/\sqrt{\bar{n}}$. This is the shot-noise limit, also known as the standard quantum limit (SQL), which we shall define as $\Delta\varphi_{\rm SQL} \equiv 1/\sqrt{\bar{n}}$, and represents the best one can do with classical-like quantum states of light, i.e. the coherent states. Even so, notice that this level of precision is obtained only for $\varphi = \pi/2$ which means that the detection of small phase shifts, as would be expected in gravitational wave detectors, would have a high degree of uncertainty. (Of course, one could compensate for this by inserting a $\pi/2$ phase-shifting element which would have the effect of replacing $\sin\varphi$ with $\cos\varphi$ in Eq. (12).) The fact that $\Delta J$ does not vanish is an indication that the quantum fluctuations of the coherent state *and* the vacuum field inputs to the first beam splitter do have the effect of limiting the precision of phase shift measurements (a coherent state has the same degree of quantum fluctuations as does the vacuum state [37]).

The precision of the phase shift measurement is not the only parameter of concern. We need also to consider the signal-to-noise ratio (SNR). In the present case this is $\text{SNR} = |\langle \hat{J} \rangle| / (\Delta \hat{J})$ or $\text{SNR} = \sqrt{\bar{n}} |\cos \varphi|$. We note that the sensitivity is low (i.e. $\Delta \varphi$ large) when the SNR is high and *vice-versa*. Ideally, of course, we'd like high sensitivity *and* high SNR.

For the sake of completeness, suppose we consider as input to the MZI, a single-photon state rather than a coherent state, the former being the apotheosis of a non-classical state of the field. We then have as the input states $|1\rangle_a |0\rangle_b$, and the first beam splitter causes the transformation

$$|1\rangle_a |0\rangle_b \xrightarrow{BS_1} \frac{1}{\sqrt{2}} \left[ |1\rangle_a |0\rangle_b + i |0\rangle_a |1\rangle_b \right] \tag{13}$$

where we notice the phase factor *i* picked up by the term containing the reflected photon. We see a difference in comparison with the coherent state case. The state on the right hand side of Eq. (13) involves entanglement.

Actually, the claim that the one photon state of Eq. (13) is entangled has given rise to a bit of controversy because superficially it appears that a single particle, a photon in this case, is somehow entangled. How can a single particle, or a single degree of freedom, be in an entangled state? Well, it can't. The single photon is not entangled, rather it is the output *a* and *b beam modes* of the beam splitter that are entangled. It is the two modes are the "particles" and the photon is an excitation whose mode of occupation is objectively indefinite. The debate over entanglement was discussed and clarified, recently by van Enk [40]. The fact that we sometimes speak of single-photon entangled states, or for that matter of *N*-photon entangled states, i.e. the N00N states (see below), is due to imprecise language. What we really mean is that, factually, the location (mode) of the *N* photons is objectively indefinite.

To return to single-photon interferometry, the phase shift operator modifies the right hand side of Eq. (13) to

$$\frac{1}{\sqrt{2}}\left[\left|1\right\rangle_a\left|0\right\rangle_b + ie^{i\varphi}\left|0\right\rangle_a\left|1\right\rangle_b\right]. \tag{14}$$

At the second beam splitter, this state transforms into the output state

$$\left|\text{out}\right\rangle = \frac{1}{2}\left[\left(1-e^{i\varphi}\right)\left|1\right\rangle_a\left|0\right\rangle_b + i\left(1+e^{i\varphi}\right)\left|0\right\rangle_a\left|1\right\rangle_b\right], \tag{15}$$

where we have applied Eq. (13) *mutatis mutandis*. We find that $\langle\hat{J}\rangle = \cos\varphi$, thus demonstrating the possibility of interference at the level of a single photon, a demonstration that was performed in the laboratory for the first time by Grangier *et al*. [41]. Note that this result agrees with the coherent state results for the case $|\alpha|^2 = 1$ and has the same dependence on the phase for all other values of $|\alpha|^2$, i.e. single photon and coherent state interferometry have the same degree of resolution though they do not have the same degree of precession. The fact that the result for $\langle\hat{J}\rangle$ in the single photon case has the same appearance as for the coherent state case with $|\alpha|^2 = 1$ should not be taken to mean that there is any direct relationship between coherent states and single-photon states: a coherent state is always a coherent state no matter the size of $|\alpha|$ and in no way approximates a state containing any definite number of photons. Furthermore, as we have shown, upon beam splitting, coherent states become a product of coherent states (they are factorized) whereas number states create entanglement.

To increase the sensitivity of the MZI with input coherent light one could apparently just increase the intensity of the light, that is, increase $\bar{n} = |\alpha|^2$. However, as pointed out by Caves [17], increasing the input laser power increases the fluctuations of the radiation pressure forces on the mirrors of the interferometer leading to errors. Thus there is the desire to enhance the sensitivity without increasing other sources of error by not arbitrarily increasing the intensity of the light. This can be done using states of light that are "nonclassical." The condition under which a state of light is non-classical is properly understood as a condition on a certain quasi-probability phase space distribution known as the *P*-function, that the function can be negative in some region of phase space or more singular than a delta function [42]. However, for our purposes we can take the following criteria for a pure single-mode field state $|\psi\rangle$ to be nonclassical: If the field is mixed with a vacuum at a beam splitter, and if the output modes are entangled, the state $|\psi\rangle$ is non-classical. As we've already shown, an input coherent state along with a vacuum in the other input mode results in products of coherent states. The coherent state is the only pure state that does not become entangled upon beam splitting [43]. Thus if we try using other input states to perform interferometry, we can expect entanglement of the beam modes inside the interferometer. Caves [17] showed that by simultaneously injecting into the input ports of the first beam splitter of the interferometer coherent and squeezed states, the latter injected into the port previously containing only the vacuum, one could reduce the level of the phase fluctuations to $\Delta\varphi = e^{-2r}/\sqrt{\bar{n}}$ where $r$ is the squeeze parameter and where $r > 0$. Thus the sensitivity of the device is enhanced over the best that can be done with classical-like light for the same average photon number, the standard quantum limit, $\Delta\varphi_{SQL} = 1/\sqrt{\bar{n}}$. Many other input states have been considered though we shall not list them all here.

The standard quantum limit demarks the boundary between the classical and quantum worlds, at least as they pertain to sensitivity for interferometery. Is there a limit on sensitivity imposed by quantum mechanics itself? Yes, there is. If phase shifts are represented by unitary operators of the form $\hat{U}_{PS} = \exp(i\varphi\,\hat{n})$, where $\hat{n}$ is the photon number operator of the relevant mode, then the uncertainty in the measurement of the phase shift is given by $\Delta\varphi_{HL} = 1/\bar{n}$, this being known, as mentioned in the introduction, as the Heisenberg limit (HL). We note an improvement over the standard quantum limit by a factor of $1/\sqrt{\bar{n}}$. How can this be achieved?

### III. Maximally entangled number states: the N00N states

We consider now a somewhat different Mach-Zehnder interferometer configuration wherein the first beam splitter is replaced by some kind of device, a "magic" beam splitter, if you will, as pictured in Fig. 4, which for the inputs $|N\rangle_a|0\rangle_b$ produces the output state

$$|\psi_N\rangle = \frac{1}{\sqrt{2}}\left[|N\rangle_a|0\rangle_b + e^{i\Phi_N}|0\rangle_a|N\rangle_b\right], \tag{16}$$

where $\Phi_N$ is some phase that may depend on $N$ and will generally depend on the method by which the states are manufactured. These are the so-called N00N states which Jon Dowling and collaborators [44-47] have been advocating for use in supersensitive, high precision, quantum metrology and in quantum sensing. The origin of the moniker N00N state is obvious, though such states are also known as maximally path-entangled number states. They are maximally entangled in the sense that the state consists of a superposition with components wherein all $N$ photons are in one mode (path) or the other. The state is entangled because in no basis can it be written as a product state of the two separate modes; i.e. $|\psi_N\rangle \neq |\psi_a\rangle \otimes |\psi_b\rangle$. According to the Copenhagen interpretation of quantum mechanics, the superposition of Eq. (16) means that the *path* of the *definite* number of photons $N$ is *objectively indefinite*. Applying the above phase shift operator to this state we obtain

$$|\psi_N(\varphi)\rangle := \hat{U}_{\text{PS}}|\psi_N\rangle = \frac{1}{\sqrt{2}}\left[|N\rangle_a|0\rangle_b + e^{i(N\varphi+\Phi_N)}|0\rangle_a|N\rangle_b\right] \quad (17)$$

where we notice the appearance of $N\varphi$ in the relative phase factor between the components of the superposition. That combination is important as it will result in a super-resolved signal, one that oscillates with $N\varphi$. As discussed in the Appendix, the output beam splitter state when the state of Eq. (17) is the input is

$$|\text{out}, N\rangle = \frac{1}{\sqrt{2^{N+1}N!}}\left[\left(\hat{a}^\dagger + i\hat{b}^\dagger\right)^N + e^{i(N\varphi+\Phi_N)}\left(\hat{b}^\dagger + i\hat{a}^\dagger\right)^N\right]|0\rangle_a|0\rangle_b. \quad (18)$$

For $N=1$ we obtain

$$|\text{out}, 1\rangle = \frac{1}{2}\left[\left(1+ie^{i(\varphi+\Phi_1)}\right)|1\rangle_a|0\rangle_1 + \left(i+e^{i(\varphi+\Phi_1)}\right)|0\rangle_a|1\rangle_1\right] \quad (19)$$

If we calculate $\langle \hat{J} \rangle = \langle \hat{n}_b - \hat{n}_a \rangle$ we find that $\langle \hat{J} \rangle = \sin(\varphi + \Phi_1)$. If we should choose $\Phi_1 = \pi/2$ phase shift, this has the same dependence on $\varphi$ as in the one photon case discussed above. For the case $N = 2$ we obtain

$$|\text{out}, 2\rangle = \frac{1}{4}\left[ \sqrt{2}\left(1 - e^{i(2\varphi+\Phi_2)}\right)|2\rangle_a |0\rangle_b + 2i\left(1 + e^{i(2\varphi+\Phi_2)}\right)|1\rangle_a |1\rangle_b \right. \\ \left. + \sqrt{2}\left(e^{i(2\varphi+\Phi_2)} - 1\right)|0\rangle_a |2\rangle_b \right]. \tag{20}$$

But in this case we find that $\langle \hat{J} \rangle = 0$. In fact, we find that $\langle \hat{n}_a \rangle = 1 = \langle \hat{n}_b \rangle$ so that even the expectation values of the number of each mode do not depend on the phase shift $\varphi$. Note, though, that $\langle \hat{n}_a + \hat{n}_b \rangle = 2$ as we would expect. It turns out that for all N00N states with $N > 1$ we find that $\langle \hat{J} \rangle = 0$ and thus no information on the phase can be obtained by this method.

So how can one extract phase-shift information from N00N states? Some years ago, the Dowling collaboration [44-46] introduced the Hermitian operator $\hat{\Sigma}_N = |N, 0\rangle\langle 0, N| + |0, N\rangle\langle N, 0|$, where the compact notation $|N, 0\rangle := |N\rangle_a |0\rangle_b$ is used. The expectation value of this operator with respect to the state of Eq. (17) is $\langle \hat{\Sigma}_N \rangle = \cos(N\varphi)$, a signal which depends on the phase shift in the combination $N\varphi$ and is thus super-resolved, the oscillation period of the interference fringes being $N$ times shorter that for fringes due to a single photon. Super-resolution is necessary for the detection of small phase shifts, but we also need super-precision as well to reduce the noise of the measurement and that's what quantum mechanics can do for us. To see this, the error propagation calculus this time leads to a phase uncertainty given by

$$(\Delta \varphi)^2 = \frac{\langle \hat{\Sigma}_N^2 \rangle - \langle \hat{\Sigma}_N \rangle^2}{\left(\partial \langle \hat{\Sigma}_N \rangle / \partial \varphi\right)^2} = \frac{1 - \cos^2(N\varphi)}{N^2 \sin^2(N\varphi)} = \frac{1}{N^2}, \quad (21)$$

where we have used the fact that $\langle \hat{\Sigma}_N^2 \rangle = 1.$ This result, which is independent of the phase shift $\varphi$, scales as $1/N$, an improvement by a factor of $1/\sqrt{N}$ over the SQL $\Delta \varphi_{\text{SQL}} = 1/\sqrt{N}$ for $N$ photons passing through the interferometer. This is the Heisenberg limit $\Delta \varphi_{\text{HL}} = 1/N$.

This is all well and good, but there's the question of the physicality of the operator $\hat{\Sigma}_N$. Though the operator is Hermitian, it is not evident as to what physical observable it represents. Notice, too, that the operator was applied directly to the N00N state and thus does not explicitly describe the effect of the beam splitter and whatever measurements are to be performed at its output. We require some kind of measurement after the beam splitter which is dependent upon the phase shift so as to give super-resolution and which, we hope, has Heisenberg limited sensitivity. Again, for these states, the standard measurement of subtraction of the output intensities vanishes and is therefore unusable. Our proposal [20-24] for a phase-sensitive measurement is the following: perform photon number parity measurements on just one of the output modes of the beam splitter. Let's take this to be the $b$ mode. The parity operator for this mode is $\hat{\Pi}_b = (-1)^{\hat{n}_a} = \exp(i\pi \hat{n}_b)$. Measuring the parity is equivalent to measuring all the moments of the operator $\hat{n}_b = \hat{b}^\dagger \hat{b}.$ We shall address the issue of how parity can be measured in section **VI**.

As already mentioned, the idea of using parity measurements for the purpose of Heisenberg-limited interferometry originates from a proposal by Bollinger *et al.* [30] in the context of Heisenberg-limited Ramsey-like [31] spectroscopy for a collection of trapped two-level ions. In this case the parity measurements refer to excited or ground state populations of the ions. Their scheme has been implemented for a small number of trapped ions. Our use of parity is an optical adaptation of the proposal of Bollinger *et al.* [30].

To see how the optical parity operator works, we consider the state $|\text{out}, 2\rangle$ and note the sign change in the middle term upon application of the mode-*b* parity operator:

$$\hat{\Pi}_b |\text{out}, 2\rangle = \frac{1}{4}\Big[\sqrt{2}\big(1 - e^{i(2\varphi + \Phi_2)}\big)|2\rangle_a |0\rangle_b \\ -2i\big(1 + e^{i(2\varphi + \Phi_2)}\big)|1\rangle_a |1\rangle_b + \sqrt{2}\big(e^{i(2\varphi + \Phi_2)} - 1\big)|0\rangle_a |2\rangle_b \Big]. \tag{22}$$

Thus we find that

$$\langle \text{out}, 2 | \hat{\Pi}_b | \text{out}, 2 \rangle = -\cos(2\varphi + \Phi_2) \tag{23}$$

which has the desired dependence on $2\varphi$. For the general case of N00N states with arbitrary *N* we have

$$\langle \hat{\Pi}_b \rangle = \frac{i^N}{2}\Big[e^{i(N\varphi + \Phi_N)} + (-1)^N e^{-i(N\varphi + \Phi_N)}\Big] \tag{24}$$

or

$$\langle \hat{\Pi}_b \rangle = \begin{cases} (-1)^{N/2} \cos(N\varphi + \Phi_N), & N \text{ even,} \\ (-1)^{(N+1)/2} \sin(N\varphi + \Phi_N), & N \text{ odd,} \end{cases} \tag{25}$$

Obviously, these are highly oscillatory functions of $\varphi$, that is, the signal displays super-resolution that scales with *N*. Examples are given in Fig. 5. It is clear that the quantity $\left|\partial\langle\hat{\Pi}_b\rangle/\partial\varphi\right|$ will be quite large and will be larger for large $N$. In fact, the phase uncertainty is Heisenberg-limited:

$$(\Delta\varphi)^2 = \frac{1-\langle\hat{\Pi}_b\rangle^2}{\left(\partial\langle\hat{\Pi}_b\rangle/\partial\varphi\right)^2} = \frac{1}{N^2}, \tag{26}$$

or $\Delta\varphi = 1/N$. We have used the fact that $\hat{\Pi}_b^2 = \hat{\mathbf{I}}$, where $\hat{\mathbf{I}}$ is the identity operator. Note that the expectation values of the parity operator in Eq. (25) are not equivalent to the expectation value $\langle\hat{\Sigma}_N\rangle = \cos(N\varphi)$ obtained from the phase-shifted N00N state of Eq. (17). There really is no reason to expect equivalence as the operator $\hat{\Sigma}_N$ is not directly connected to an observable.

For the use of N00N states within the parity measurement scheme, the SNR is found to be

$$\text{SNR} = \begin{cases} \left|\dfrac{\cos(N\varphi+\Phi_N)}{\sin(N\varphi+\Phi_N)}\right|, & N \text{ even} \\ \left|\dfrac{\sin(N\varphi+\Phi_N)}{\cos(N\varphi+\Phi_N)}\right|, & N \text{ odd.} \end{cases} \tag{27}$$

Evidently, the SNR can be large in the vicinity of $\varphi \approx 0$ for $N$ even if $\Phi_N = 0$ or for $N$ odd for $\Phi_N = \pi/2$.

On the experimental side of using N00N states, some years ago, Rarity *et al.* [48] observed super-resolved interference in a N00N state of two photons, and later Kuzmich and Mandel [49] performed sub-shot noise interferometric measurements with a state of the same type. More recently, super-resolved phase measurements have been carried out using N00N states for $N = 3$ and $N = 4$ photons [50, 51]. However, the phase-shift measurements for these experiments were not sub-shot noise and the SNR was low. None of these experiments [48-51] involved the use of parity measurements. The signals for all of them were obtained from coincidence counting on the output beams of a beam splitter.

In general, it is difficult to generate the number states $|N\rangle$ required for the N00N states, especially for high $N$, and difficult to generate the N00N state out of the number states. The latter generally requires a nonlinear process. In response to the problem of generating large photon number state, we have studied [20, 21, 23] the prospect of using maximally entangled coherent states of the generic (unnormalized) form $|\alpha\rangle_a|0\rangle_b + e^{i\Phi}|0\rangle_a|\alpha e^{i\theta}\rangle_b$, as the coherent states themselves are relatively easy to generate. The maximally entangled coherent states are just superpositions of the N00N states. We have found for small phase shifts $\varphi$ that one can achieve the phase uncertainty $\Delta\varphi \approx 1/\bar{n}$ where $\bar{n} = |\alpha|^2$. Of course, one still needs a nonlinear interaction to produce a maximally entangled coherent state out of an input coherent state $|\alpha\rangle$. A discussion of these issues is beyond the scope of this article.

## IV. Interferometry with twin Fock states

In 1993, Holland and Burnett [57] proposed another approach to Heisenberg-limited interferometry using a normal MZI (beam splitters at the input and the output) but where identical photon number states (twin Fock states) are simultaneously fed into the first beam splitter. That is, the input state is $|N\rangle_a|N\rangle_b$ as indicated in Fig. 6. Holland and Burnett [57] argued that the phase uncertainty should approximately be $\Delta\varphi \approx 1/(2N)$, where we now appropriately have $2N$ in the denominator as that is the total number of photons passing through the interferometer. But their scheme suffers from the same problem as does the N00N state: the expectation value of $\hat{J} = \hat{n}_b - \hat{n}_a$ vanishes, $\langle \hat{J} \rangle = 0$, and so there is no dependence on the phase-shift. This led us to the idea of applying the parity operator method to Holland-Burnett scheme [25, 26]. Before we discuss that application, let's first see what a beam splitter does to the input state $|N\rangle_a |N\rangle_b$.

For later convenience, we are going to assume that the first beam splitter of the MZI of Fig. 6 is constructed so that it *does not* give rise to a $\pi/2$ phase shift in the reflected beam. The beam splitter transformations are now

$$\hat{a}' = \frac{1}{\sqrt{2}}(\hat{a} + \hat{b}), \quad \hat{b}' = \frac{1}{\sqrt{2}}(\hat{b} - \hat{a}). \tag{28}$$

Using the same techniques as in the Appendix but now with the transformation of Eq. (28), starting with $|N\rangle_a |N\rangle_b = \hat{a}^{\dagger N} \hat{b}^{\dagger N} |0\rangle_a |0\rangle_b / N!$ the output state of the first beam splitter is given by

$$|\psi_{2N}\rangle = \frac{1}{\sqrt{2^{2N}} N!} \left( \hat{a}^\dagger - \hat{b}^\dagger \right)^N \left( \hat{a}^\dagger + \hat{b}^\dagger \right)^N |0\rangle_a |0\rangle_b, \tag{29}$$

where we have again dropped the primes. For $N = 1$ we obtain

$$|\psi_2\rangle = \frac{1}{\sqrt{2}} \left( |2\rangle_a |0\rangle_b - |0\rangle_a |2\rangle_b \right) \tag{30}$$

which is, of course, just a N00N state with two photons. Such a state was first generated and detected in a famous experiment performed by Hong, Ou, and Mandel in 1987 [58]. For $N = 2$ we find that

$$|\psi_4\rangle = \sqrt{\frac{3}{8}} \left( |4\rangle_a |0\rangle_b + |0\rangle_a |4\rangle_b \right) - \frac{1}{2} |2\rangle_a |2\rangle_b. \tag{31}$$

This state was first generated and detected in 1999 by Ou, Rhee, and Wang [59]. It is not a N00N state but it does *contain* a four-photon N00N. In fact, for all $N$ there will be a N00N state component of the form $|2N\rangle_a|0\rangle_b \pm |0\rangle_a|2N\rangle_b$, a fact which is strongly suggestive that the twin Fock state input approach might lead to Heisenberg-limited phase uncertainty.

For arbitrary $N$, the state just after the first beam splitter, $BS_1$, is [25, 60]

$$|\psi_{2N}\rangle = \sum_{k=0}^{N} A_k^N |2k\rangle_a |2N-2k\rangle_b \tag{32}$$

where

$$A_k^N = \frac{(-1)^{N-k}}{2^N} \left[ \binom{2k}{k} \binom{2N-2k}{N-k} \right]^{1/2}. \tag{33}$$

Including the phase shift picked up in the inside the interferometer, we have

$$|\psi_{2N}(\varphi)\rangle = \sum_{k=0}^{N} A_k^N e^{i\varphi(2N-2k)} |2k\rangle_a |2N-2k\rangle_b. \tag{34}$$

Without going into the details, following through the second beam splitter and taking the expectation value of the $b$-mode parity operator of the MZI output we find that $\langle \hat{\Pi}_b \rangle = P_N[\cos(2\varphi)]$ where the $P_N(x)$ is a Legendre polynomial. In Fig. 7 we plot $\langle \hat{\Pi}_b \rangle$ versus $\varphi$ for different $N$ where we observe super-resolved oscillations scaling with $2N$. In Fig. 8 we plot the numerically obtained phase uncertainty against $2N$, along with those of the standard quantum limit and $\Delta\varphi_{SQL} = 1/\sqrt{2N}$ and the Heisenberg limit $\Delta\varphi_{HL} = 1/(2N)$, where we have taken into account the fact $2N$ photons pass through the interferometer. We see that for $\varphi = 0$ the sensitivity for the twin Fock state inputs is very close to the Heisenberg limit for all *N*. If $\varphi$ deviates too much from zero, the sensitivity is reduced over some ranges of *N* though there are also ranges which are still near the HL. Numerically we can show that the SNR for the twin Fock state case is generally very large. We point out that Kim *et al*. [61] showed that the operator $\hat{J}^2$ could be used as a phase-shift measure to achieve Heisenberg-limited phase uncertainty, but the maximum of the SNR is a mere $\sqrt{2}$.

As N00N states are difficult to generate, so are twin-Fock states. One must have photon number states of identical large photon numbers presented simultaneously at the two input ports of a beam splitter, a considerable challenge. On the other hand, there are optical processes that produce superpositions of twin-Fock states of the form $\sum_{N=0}^{\infty} C_N |N\rangle_a |N\rangle_b$. An important example of such a state is the two-mode squeezed vacuum state which is routinely produced in laboratories by the process of parametric spontaneous down-conversion [62]. The two-mode squeezed vacuum state has recently been considered for interferometry by Anisimov *et al*. [63] who have shown that the phase uncertainty obtained from such a state actually beats the Heisenberg limit, though the improvement is small generally. But there is another kind of state to be considered, namely a pair coherent state [64]. Though the pair coherent states have yet to be generated in the laboratory, they have advantages for interferometry over the two-mode squeezed vacuum states in that the latter has a thermal-like photon number distribution (which is very broad, super-Poissonian, and peaked at the vacuum states) whereas the former has a narrow (sub-Possionian) distribution peaked near the average photon, which can be quite high. For these reasons, the two-mode squeezed vacuum states do not exhibit super-resolution whereas the pair coherent states do [65].

**V. Remarks on the connection between Heisenberg-limited phase uncertainty and photon number distributions**

Why do we obtain Heisenberg-limited phase uncertainty with some input states but not others? Here we give a qualitative rationale based on the heuristic number-phase uncertainty relation $\Delta N \Delta \varphi \geq 1$.

The key seems to be the nature of the joint photon number probability distribution $P(n_1, n_2)$ inside the interferometer. Consider the joint photon number distribution of the N00N state of Eq. (16) for $N = 10$ as pictured in Fig. 9. Its only nonzero probabilities are $P(N,0) = 1/2 = P(0,N)$ indicating a large fluctuation in the number of photons in each of the modes. In fact, for each of the modes the fluctuations are clearly given by $\Delta N \sim N$ which, from the number-phase uncertainty relation leads to $\Delta \varphi \sim 1/N$, the Heisenberg limit.

For the case of the input twin-Fock states $|N\rangle_a |N\rangle_b$, using the state $|\psi_{2N}(\varphi)\rangle$ of Eq.(34), we find that the only non-zero joint probabilities are

$$P(2k, 2N-2k) = |A_k^N|^2 = \frac{1}{2^N} \binom{2k}{k}\binom{2N-2k}{N-k}, \quad \text{for } k = 0, 1, ..., N. \qquad (35)$$

As displayed in Fig. 10, the joint distribution for this state forms an anti-diagonal across the $n_1, n_2$ plane. Note that only even photon number states are populated and that the greatest probabilities are those for which $k = 0$ and $k = N$, $P(0, 2N)$ and $P(2N, 0)$ respectively, with a plateau in between. The distribution, or at least the continuous variable analog of it, is known in probability theory as the fixed-multiplicative discrete arcsine law of order $N$ [66] and for that reason the states of Eq.(34) have been dubbed the arcsine states [60]. In view of the probability distribution, the uncertainty in the photon number is $\Delta N \sim 2N$ from which we conclude that $\Delta \varphi \sim 1/(2N)$.

## VI. Parity measurement

As we have seen, photon number parity measurements enables optimal measurements of phase shifts in optical interferometry in a wide range of nonclassical input states of light and even enables optimal performance in with input classical-like states of light. By optimal we mean Heisenberg-limited phase uncertainty and high SNR. As mentioned in the Introduction, the parity operator is intimately related to the Wigner function which can be expressed as the expectation value of the displaced parity operator [7, 8]. The reconstruction of the Wigner function to determine a quantum state could be performed through the use of parity measurements. But how can we measure photon number parity?

The obvious straight forward way is to simply do a photon counting measurement and assign parity to be $+1$ if the number is even or $-1$ if the number is odd. Unfortunately, the world just isn't that simple. This requires detectors that can count photons at a resolution at the level of a single photon. Readily available photon detectors are unable to perform such precise measurements. However, there is a significant effort under way to develop low-noise photon number resolving detectors for optical and infrared photons. These include loop detectors [67], time multiplexing detectors [68, 69], avalanche photodiodes operated in a Geiger mode [70], and superconducting devices [71, 72]. The detector described in Ref. [71], for example, can resolve photon-number detections at the level of a single photon for photon numbers up to about 9 or 10. That should be enough for a proof-of-principle demonstration of our proposed measurement scheme. On the other hand, photon numbers could be measured with single-photon resolution using quantum non-demolition (QND) techniques. The basic idea was proposed some years ago [73] and has been continuously refined [74]. The technique involves the use of a nonlinear optical interaction known as the cross-Kerr interaction where a large nonlinearity is required. This requires a large third-order susceptibility $\chi^{(3)}$. Unfortunately, readily available nonlinear media have susceptibilities that are many orders of magnitudes too small. On the other hand, a technique known as electromagnetically induced transparency (EIT) [75] promises to be able to produce the required nonlinearities. Even modest enhancements of the nonlinearity can be useful for QND measurements as has been shown by Wunro *et al*. [76]

The previous paragraph discussed parity measurements obtained from first measuring photon number at high resolution. Another prospect is the direct measurement of parity with *insensitive* photo-detectors by using nonlinear optical switching devices that take incoming photons and shuttle them into one path or another depending on their parity as shown in schematic form in Fig. 27. As parity is determined by exit path, only insensitive detectors placed in those paths are required. Such a device, using a triple cross-Kerr interaction, was described by Yurke and Stoler [77]. As we showed [23], another possible device is a nonlinear interferometer, an MZI with self-Kerr media in both arms. We also showed that it is possible, in principle, to perform QND measurements of parity directly [78]. Unfortunately, all of these methods require the large nonlinear interactions which are not generally available. But we have also shown that QND measurements of photon number with weak nonlinearities are possible [79]

Lastly in this section, as we pointed out the Introduction, the parity operator and the Wigner function $W(\alpha)$ [6] are intimately related according to [7, 8] $W(\alpha) = 2\langle \hat{D}^\dagger(\alpha) \hat{\Pi} \hat{D}(\alpha) \rangle / \pi$, that is, the Wigner function is the expectation value of the displaced parity operator. Here $\hat{D}(\alpha) = \exp(\alpha \hat{a}^\dagger - \alpha \hat{a})$ is the displacement operator for a single-mode field (see Appendix). The parameter $\alpha$ is a complex number whose real and imaginary parts constitute the variables of phase space. At the origin of this phase space, $\alpha = 0$, we find that $\langle \hat{\Pi} \rangle = 2W(0)/\pi$. Thus if we know the value of the Wigner function at the origin of phase space we know the expectation value of the parity operator. A procedure has been developed called quantum tomography to construct the Wigner function of radiation fields by using a series of homodyne measurements along with the Radon transformation [80]. This method could be used to obtain the average of the parity operator without directly measuring parity. The entire Wigner function would not need to be constructed as only its value at the origin of phase space would be required. A measurement of the Wigner function at the origin has already been performed for a microwave quantized field in a cavity [81], but not, as far as we are aware, for a traveling wave optical or infrared field.

**VII. Conclusions**

In this article we have reviewed some aspects of our work on the prospect of using the photon number parity operator for the purpose quantum optical metrology. Specifically, we have discussed its application to interferometry with N00N states and with twin-Focks states injected into a beam splitter. The advantage of measuring photon parity on one of the output beams of the interferometer is that it leads to super-resolved and supersensitive (Heisenberg-limited) phase measurements, and generally leads to measurements with a high SNR. In contrast, the standard method of phase shift detection, the subtraction of output beam intensities, is insensitive to the phase shift when using N00N states or twin-Fock states. Coincidence counting techniques *are* phase-shift sensitive and are super-resolving, but they are not Heisenberg limited in precision and they have low SNR. We mention that H. Lee and collaborators [56, 82] have shown that the parity detection method has the promise of improving almost all of the various interferometry schemes that have been proposed using a variety of nonclassical states of light. In fact, they have declared it a unified detection scheme for precision phase shift measurements.

Finally we point out that very recently Caves and Shaji [83] have presented a quantum-circuit guide to many to the interferometric schemes using N00N state, maximally entangled coherent states, and other states through parity measurements.

**Acknowledgements**

CCG gratefully wishes to thank A. Benmoussa and R. A. Campos for collaborations on this work over the past decade, and for conversations with Mark Hillery, Hwang Lee, and Jon Dowling. The work has been funded by The Research Corporation, The National Science Foundation, The Army Research Office, and grants from PSC-CUNY.

**Appendix: Beam splitters acting on quantum states**

Here we briefly describe a technique useful for obtaining quantum states resulting from beam splitter transformations on certain input states. More details are given in Ref. [36].

In the following we use the being splitter transformations given already in Eqs.(4) and their inverses

$$\hat{a} = \frac{1}{\sqrt{2}}\left(\hat{a}' - i\hat{b}'\right), \quad \hat{b} = \frac{1}{\sqrt{2}}\left(\hat{b}' - i\hat{a}'\right). \tag{36}$$

It is trivial to see that input vacuum states $|0\rangle_a|0\rangle_b$ results in the output state $|0\rangle_{a'}|0\rangle_{b'}$, that is, $|0\rangle_a|0\rangle_b \to |0\rangle_{a'}|0\rangle_{b'}$ upon beam splitting. Consider now the input state $|1\rangle_a|0\rangle_b$ which we can write as $\hat{a}^\dagger|0\rangle_a|0\rangle_b$. Thus, because of the transformation property of the input vacuum state and of the operator transformations of Eq. (36), we can write

$$\begin{aligned}\hat{a}^\dagger|0\rangle_a|0\rangle_b &\to \frac{1}{\sqrt{2}}\left(\hat{a}'^\dagger + i\hat{b}'^\dagger\right)|0\rangle_{a'}|0\rangle_{b'}, \\ &= \frac{1}{\sqrt{2}}\left(|1\rangle_{a'}|0\rangle_{b'} + i|0\rangle_{a'}|1\rangle_{b'}\right),\end{aligned} \tag{37}$$

which is just the result given in Eq. (13). We can get the output state for any input state by this method. As another example, if the input state is $|1\rangle_a|1\rangle_b = \hat{a}^\dagger \hat{b}^\dagger |0\rangle_a|0\rangle_b$, we then have as the output state

$$\begin{aligned}\hat{a}^\dagger \hat{b}^\dagger |0\rangle_a|0\rangle_b &\to \frac{1}{\sqrt{2}}\left(\hat{a}'^\dagger + i\hat{b}'^\dagger\right)\frac{1}{\sqrt{2}}\left(\hat{b}'^\dagger + i\hat{a}'^\dagger\right)|0\rangle_{a'}|0\rangle_{b'}, \\ &= \frac{i}{\sqrt{2}}\left(|2\rangle_{a'}|0\rangle_{b'} + |0\rangle_{a'}|2\rangle_{b'}\right).\end{aligned} \tag{38}$$

(This differs somewhat from the result given in Eq. (30) because in that case a different type of beam splitter was assumed, one that does not give a $\pi/2$ phase shift to the reflected beam.) In general, for any input number state of the form $|N\rangle_a |M\rangle_b$, the output state can be found by expanding out the right hand side of

$$|\text{out}\rangle = \left[\frac{1}{\sqrt{2}}\left(\hat{a}'^\dagger + i\hat{b}'^\dagger\right)\right]^N \left[\frac{1}{\sqrt{2}}\left(\hat{b}'^\dagger + i\hat{a}'^\dagger\right)\right]^M |0\rangle_{a'} |0\rangle_{b'} \qquad (39)$$

As a last example, we consider the case of an input coherent state in the *a* beam with a vacuum in beam *b*: $|\alpha\rangle_a |0\rangle_b$. As is well known, the coherent states can be defined as right eigenstates of the annihilation operator, $\hat{a}|\alpha\rangle = \alpha|\alpha\rangle$, but they can also be defined as displaced vacuum states according to $|\alpha\rangle = \hat{D}(\alpha)|0\rangle$ where $\hat{D}(\alpha) = \exp(\alpha\hat{a}^\dagger - \alpha^*\hat{a})$ is the displacement operator. Writing the input state as $\hat{D}(\alpha)|0\rangle_a |0\rangle_b$ and setting $\hat{a} = (\hat{a}' - i\hat{b}')/\sqrt{2}$ and $\hat{a}^\dagger = (\hat{a}'^\dagger + i\hat{b}'^\dagger)/\sqrt{2}$ into the displacement operator we find

$$\hat{D}_a(\alpha)|0\rangle_a |0\rangle_b \to \hat{D}_{a'}\left(\frac{\alpha}{\sqrt{2}}\right) \hat{D}_{b'}\left(\frac{i\alpha}{\sqrt{2}}\right) |0\rangle_{a'} |0\rangle_{b'} = \left|\frac{\alpha}{\sqrt{2}}\right\rangle_{a'} \left|\frac{i\alpha}{\sqrt{2}}\right\rangle_{b'}, \qquad (40)$$

in agreement with what we obtained from a heuristc approach in section II above.

A more sophisticated approach to beam splitting, using Lie group and algebraic methods, can be found in Ref. [60].

**References**

[1] W.M. Gibson and B.R. Holland, *Symmetry Principles in Elementary Particle Physics*, Cambridge University Press, Cambridge, 1976.

[2] H. Frauenfelder, E. M. Henley, *Subatomic Physics*, 2[nd] Edition, Prentiss Hall, Englewood Cliffs, 1991.

**Figure captions.**

Fig. 1 Diagram of a Mach-Zehnder interferometer with an input coherent and vacuum states and photon number detection on the two output beams. The wedge in the clockwise beam represents the relative phase shift between the two beams.

Fig. 2 A 50:50 beam splitter. The reflected beam picks up a $\pi/2$ phase shift.

Fig. 3 Plot of $\langle \hat{J} \rangle$ versus $\varphi$ for input coherent states of different $\bar{n}$. The line with the squares is for $\bar{n}=5$, the line with the dots is for $\bar{n}=10$, and the line the triangles is for $\bar{n}=20$.

Fig. 4 A Mach-Zehnder-tpye interferometer where the first beam splitter has been replaced by a "magic" beam splitter, a device that can make N00N states.

Fig. 5 For N00N states, plots of the expectation value of the output $b$-mode parity against $\varphi$ for two different values of $N$. The solid line is for $N=4$ and the dashed line for $N=30$.

Fig. 6 Diagram of a Mach-Zehnder interferometer with twin-Fock state inputs and parity measurements made on the output $b$-mode.

Fig. 7 For twin-Fock state inputs, plots of the expectation value of the output $b$-mode parity against $\varphi$ for different values of $N$. The solid line is for $N=2$ and the dotted line is for $N=15$.

Fig. 8 For input twin-Fock states, plots of the phase uncertainty $\Delta\varphi$ versus $2N$ for $\varphi=0.0001$ (the squares) and for $\varphi=0.05$ (the dots). The dot-dashed line is the standard quantum limit and the dotted line is the Heisenberg limit.

Fig. 9 Plot of the joint photon number probability distribution $P(n_1,n_2)$ versus $n_1$ and $n_2$ for a N00N state with $N=10$.

Fig. 10 Plot of the joint photon number probability distribution $P(n_1,n_2)$ versus $n_1$ and $n_2$ for an arcsine state obtained from twin-Fock state inputs of $N=10$ photons each.

Fig. 1.

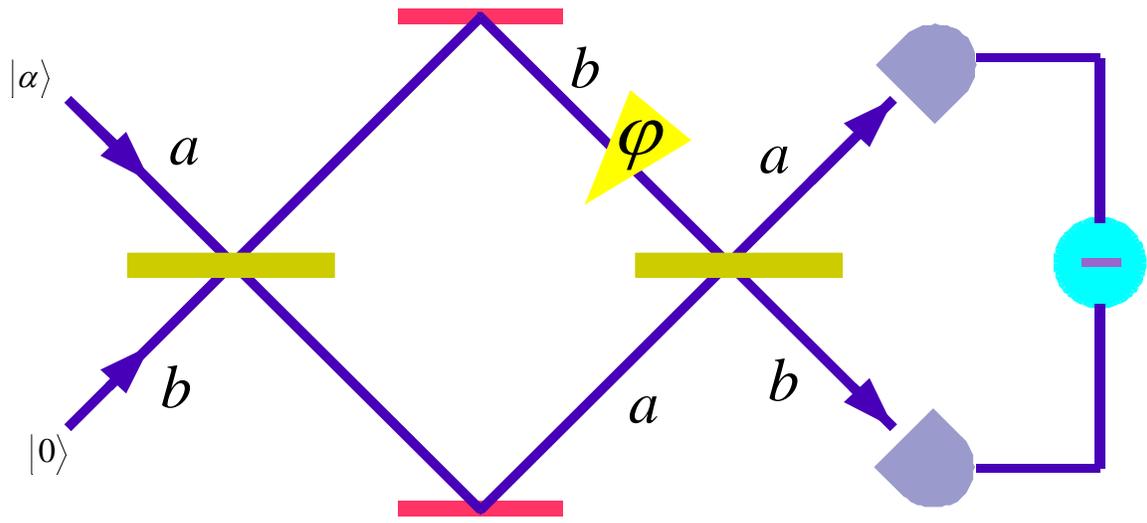

Fig. 2.

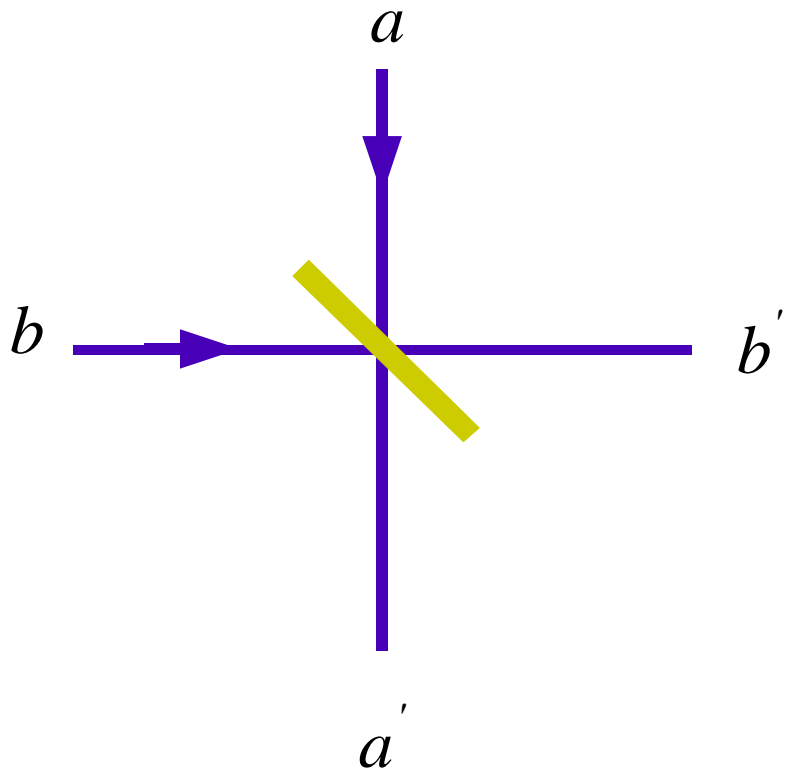

Fig. 3.

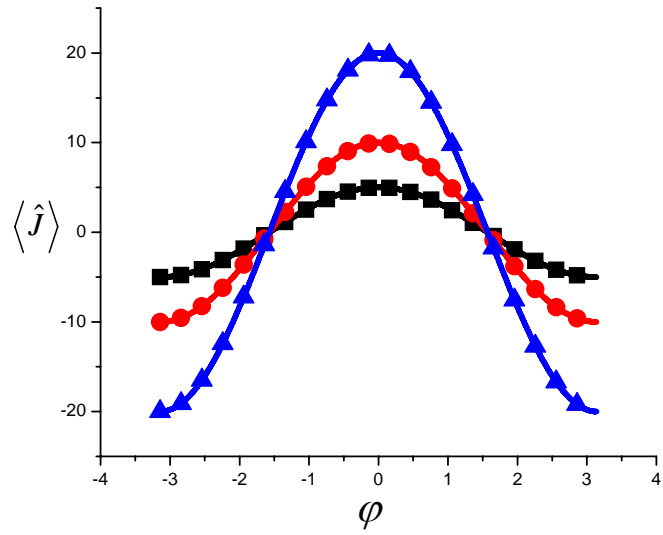

Fig. 4.

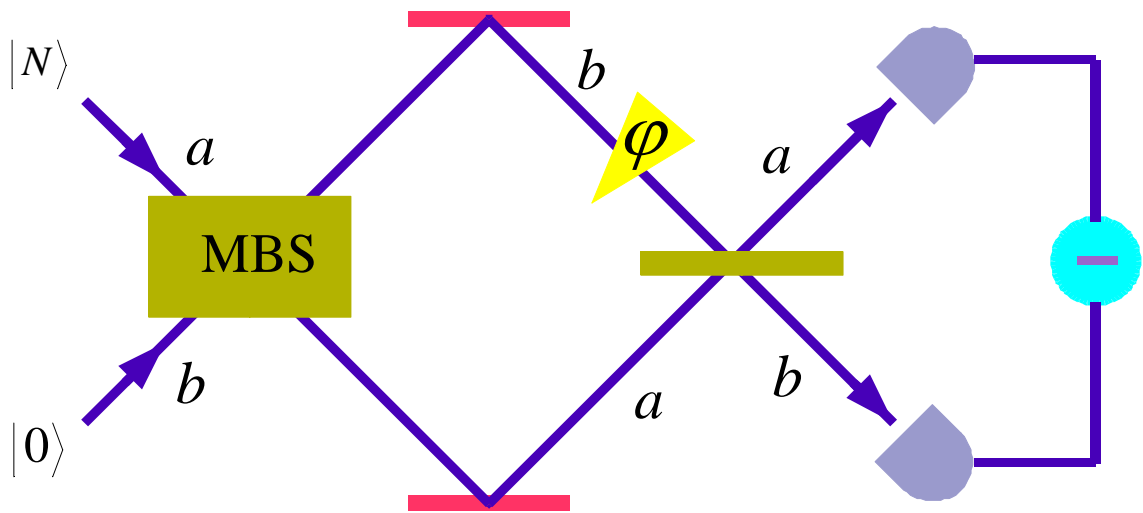

Fig. 5.

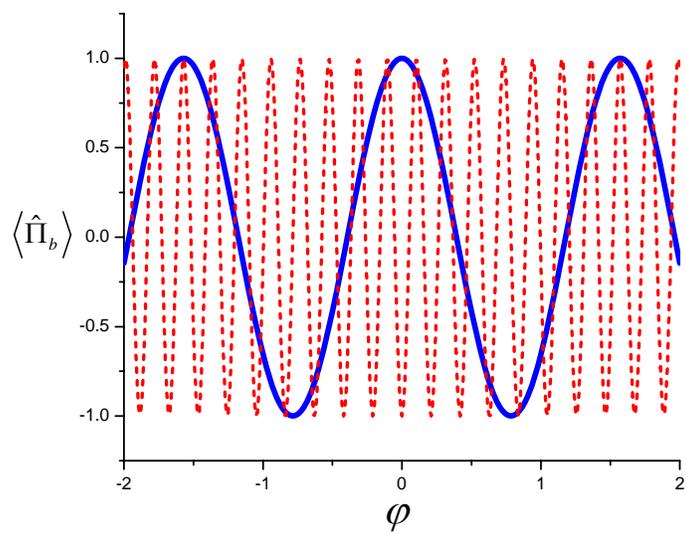

Fig. 6.

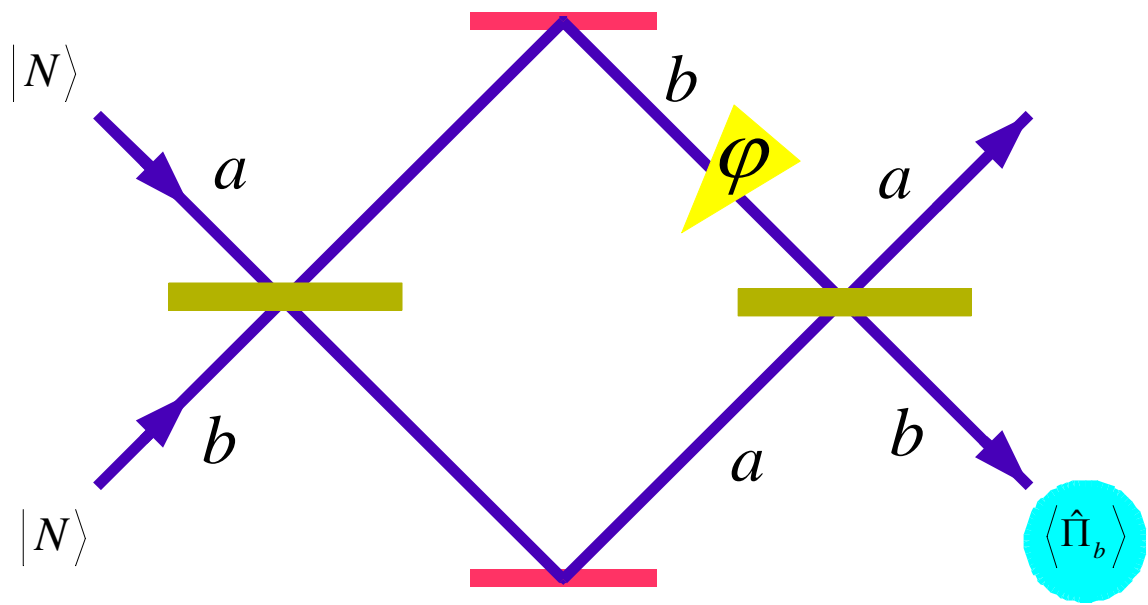

Fig. 7.

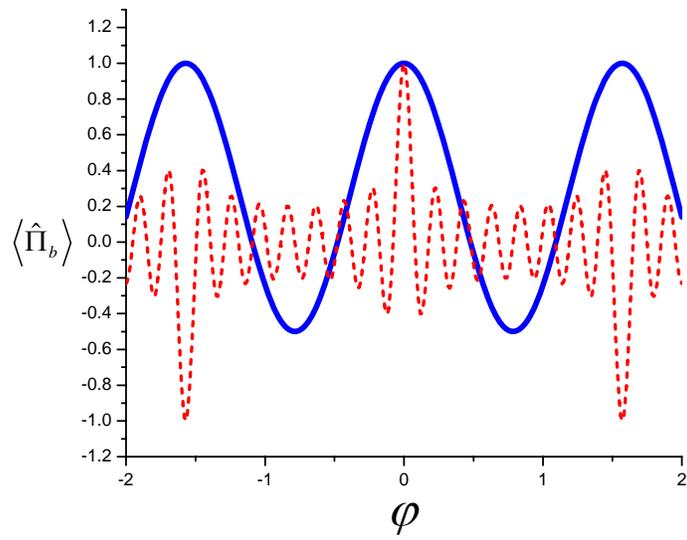

Fig. 8.

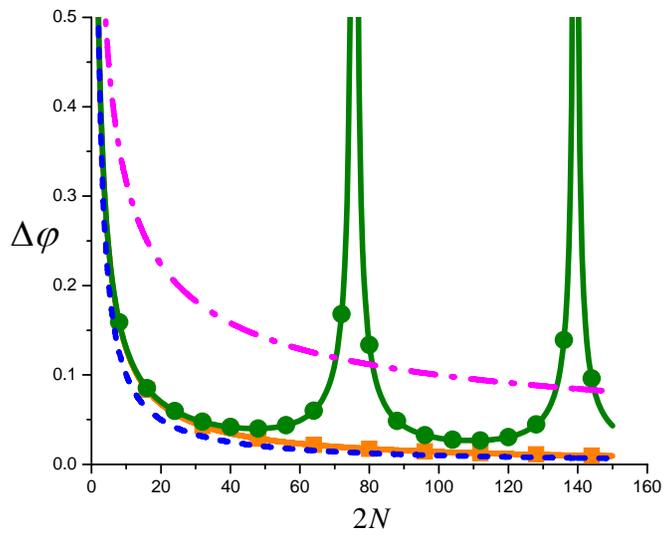

Fig. 9.

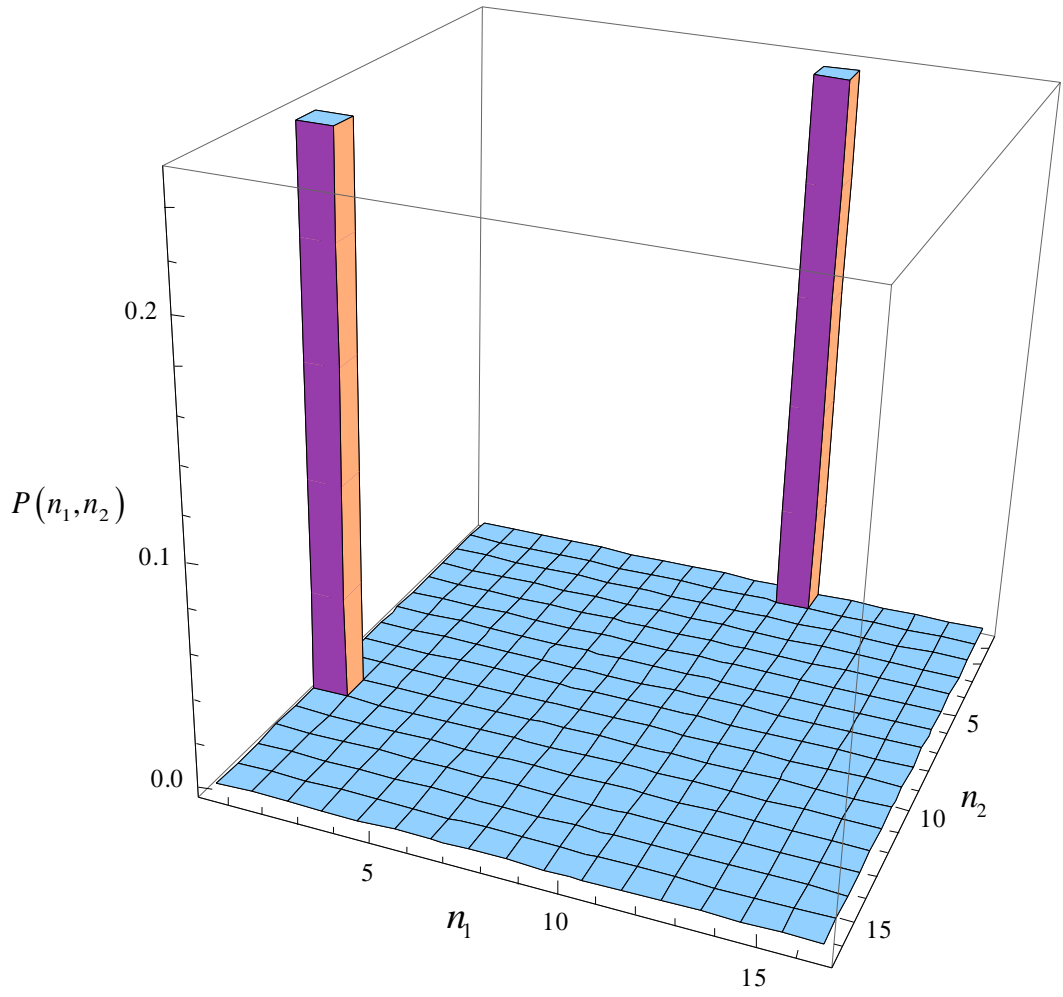

Fig. 10.

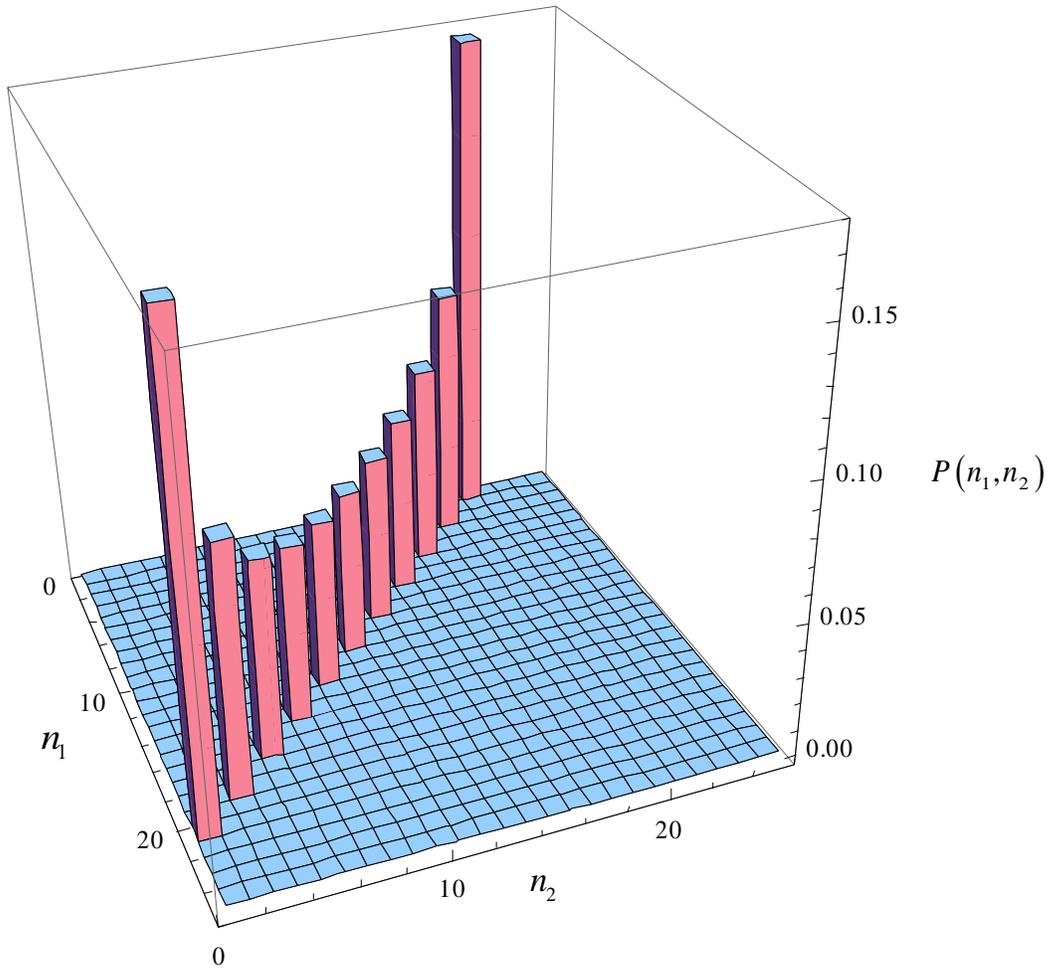